\documentclass[preprint,12pt,authoryear]{elsarticle}

\usepackage{amssymb}
\usepackage{amsmath}

\usepackage{url}
\usepackage{multirow}
\usepackage{makecell}
\usepackage{booktabs}

\journal{Computers and Geosciences}

\begin{document}

\begin{frontmatter}



\title{Image-based modelling of elastic properties using unresolved rock images} 


\author[label1]{Rui Li} 
\author[label1]{Yi Yang}
\author[label1]{Wenbo Zhan}
\author[label2]{Jianhui Yang}
\author[label1,label3]{Yingfang Zhou\corref{cor1}}\ead{yingfang.zhou@abdn.ac.uk}

\cortext[cor1]{Corresponding author}
\affiliation[label1]{organization={School of Engineering, University of Aberdeen},
            postcode={AB24 3FX},
            country={UK}}
            
\affiliation[label2]{organization={Geoscience Research Centre, Total E\&P UK Limited},
            country={UK}}

\affiliation[label3]{organization={School of Energy Resources, China University of Geosciences},
            city={BeiJing},
            postcode={100083},
            country={China}}

\begin{abstract}
The trade-off between image resolution and model field-of-view has long been a limitation for numerical simulations in digital rock models.
A significant amount of sub-resolution pore space cannot be captured in the unresolved digital rock images, which hinders the accuracy of numerical simulations, especially those predicting the rock effective elastic properties.
This work uses paired digital rock images at multiple resolutions to investigate the sub-resolution solid and pore fraction distributions.
It demonstrates that the cumulative Beta distribution function can effectively represent the solid and pore fractions in unresolved rock images.
Based on this finding, we propose a novel methodology to predict the sub-resolution pore fractions.
Compared to the pore fractions extracted from paired Bentheimer sandstone images at resolutions of 2, 6, and 18$\mu m$, the proposed method yields double-weighted mean absolute percentage errors (WWMAPE) of 3.67\% (6$\mu m$) and 13.78\% (18$\mu m$), significantly lower than the SLOT technique's errors of 24.55\% (6$\mu m$) and 59.33\% (18$\mu m$).
By incorporating the modelled solid and pore fractions with effective medium theory, this work achieves improved predictions of effective elastic moduli and wave velocities across various rock types. 
This method has significant implications for conducting enhanced simulations of geophysical properties in unresolved rock models.
\end{abstract}


\begin{highlights}
\item The cumulative Beta distribution function effectively represents the sub-resolution pore fraction in unresolved images of various rock samples.
\item The simulated rock elastic properties of the unresolved rock images agree well with documented experimental results.
\item The methodology has potential for facilitating more accurate numerical simulations in unresolved rock models.
\end{highlights}


\end{frontmatter}

\section{Introduction}
Rock elastic moduli and sound wave velocities are fundamental parameters for engineering geology applications, such as underground structure design and seismic interpretation \citep{GONZALEZDEVALLEJO2003273, SOW2017133, LI2020105262,  ZHAO2021106138, FORTE2021106194}.
In general, these properties are determined through laboratory testing of core samples \citep{SONG2004293}.
However, this approach suffers from substantial time and financial investments, as well as limited availability of suitable core samples \citep{KIM201268}.
Moreover, some testing procedures inevitably destruct samples that prevents sample reuse \citep{PAPPALARDO2022106829}.
Therefore, numerical modelling techniques, taking advantage of imaging techniques (e.g., micro-CT), has emerged as an effective non-destructive approach for rock elastic properties predictions \citep{Andra2013a, FarhanaFaisal2019, Li2022, Goldfarb2022, Alqahtani2022, Wildenschild2013, Al-Kharusi2007}. 
The modelling workflow typically comprises three key steps \citep{Andra2013a}.
The first step involves scanning rock samples using X-ray computed tomography, resulting in 2-D rock images stacks that capture pore-scale features. 
These images consist of voxels with intensity values (or CT numbers) proportional to material density.
The second step entails constructing rock models based on these digital images, which requires the use of image segmentation methods to identify different phases. 
In a conventional threshold-based segmentation approach for bimodal rock samples, a global intensity threshold is selected. 
Voxels with intensity values below this threshold are considered as the pore phase, while those above it are identified as the solid phase.
Following the image segmentation process, the third step involves computing various geophysical properties using different numerical solvers \citep{Saxena2017, Saenger2016, Dvorkin2011, Blunt2013}.
This technique allows for the capture of the intricate geometry of rocks and the estimation of their geophysical properties without causing any damage to the original rock samples. 
Consequently, it serves as an alternative method for investigating rock properties, supplementing traditional laboratory experiments and mathematical models.

However, the accuracy of numerical simulation technique in terms of rock elastic properties prediction has remained a longstanding challenge over the past decades. 
Many studies have reported that the predicted elastic properties are systematically overestimated when compared to experiment measurements \citep{Madonna2012, Saxena2017a, Arns2002a, Devarapalli2017, Goldfarb2022, Ikeda2020}.
This inconsistency arises from various uncertainties, including discrepancies in sample scale compared to laboratory measurements, limitations of numerical solvers, image segmentation methods, inadequate identification of mineralogy and input of elastic moduli, and the restricted image resolution \citep{Saxena2017a, Andra2013, FarhanaFaisal2019}.
Among these sources of uncertainty, limited image resolution stands out as a primary reason for poor simulation results \citep{FarhanaFaisal2019, Knackstedt2009}. 
A coarse image resolution is unable to resolve fine rock features, such as micropores and grain interfaces, which are crucial for effectively modelling rock mechanical properties. 
While extremely high image resolution can capture more of these fine features, its use is often avoided in many studies due to the substantial computational time required and limitations in the scale of views, particularly for highly heterogeneous carbonate rocks.
Furthermore, some rocks, such as carbonate rocks, contain pores with sizes down to the nanometer scale, which makes it nearly impossible to find a single resolution that resolves all rock features \citep{Jouini2015}.
\citet{Jouini2015} proposed a method to estimate the effective elastic properties of carbonate rock samples using rock images at multiple resolutions.
First, they scanned carbonate rock core plugs at a resolution of $19 \mu m$. 
They then scanned some microplugs at higher resolutions, ranging between 0.3 and $2 \mu m$. 
The effective elastic properties of the microplugs were calculated and extrapolated to the entire unresolved phase. 
This approach allowed them to compute the effective elastic properties of the whole core plug.
However, the simulated elastic properties still significantly exceed the experimental results, with the highest error exceeding 150\% for the tested carbonate rock samples. 

The primary challenge associated with using unresolved images is the partial-volume effect, which arises from low image resolution and blurred phase boundaries \citep{Schluter2014, Korneev2018, Gerke2015}. 
Partial-volume voxels encompass both pore and solid phases and are situated around phase interfaces, creating a broad, blurred phase edge. 
This blurred edge can be wider than many fine pore structures, causing these pore structures to be completely encompassed by voxels within the blurry phase. 
Extracting these pore structures necessitates a very high intensity threshold in conventional global thresholding methods. 
Consequently, these fine features may be lost in the reconstructed models, resulting in solid grains merging together to form a compact solid framework across the domain, which significantly increases the stiffness of the model.
Enhancing the quality of reconstructed rock models is thus of paramount importance in improving the accuracy of simulation results. 
This underscores the need for a comprehensive consideration of the partial-volume effect in image segmentation methods.

Efforts have been made to address the discrepancy between modelled mechanical properties and experiment measurements. 
\citet{Saenger2016} suggested that using a watershed segmentation method to extract the grain boundaries.
By assigning moderate elastic moduli to these boundary voxels, it becomes possible to accurately estimate the effective elastic properties of the model.
Developing rock models with multiple resolutions can also improve the accuracy of elastic property simulations \citep{FarhanaFaisal2019,Jouini2015}. 
This technique involves scanning rock samples at two resolutions. 
The lower resolution captures the macro rock features and provides a sufficiently large field of view. 
The 'solid phase' captured at the low resolution is not purely solid but a combination of pore and solid phases. 
Therefore, higher resolution images are needed to better characterise this 'solid phase' and determine its effective elastic properties. 
Subsequently, these corrected elastic properties are extrapolated to the solid phase in the low-resolution model. 
This correction leads to an improved prediction of the effective elastic properties.
Additionally, a 'segmentation-less' method has been proposed for calculating the effective elastic moduli and has been successfully tested on Berea sandstones \citep{Goldfarb2022, Ikeda2020}. 
This method selects intensity targets (either actual or pseudo targets) to calibrate the relationship between intensity values of partial-volume voxels and their density. 
This calibration, in turn, determines the relationship between image intensity values and pore fraction within each partial-volume voxel.
Using the pore fraction, the effective medium theory is adopted to assign the bulk and shear moduli to each partial-volume voxel. 
Ultimately, the effective elastic properties of the entire model are computed.
Notably, the modelled elastic moduli for the Berea sandstone sample align well with experiment measurements. 
However, the simple relationship between image intensity values and pore fraction adopted in this method is either linear or power-law, which is based on hypotheses and lacks solid evidence.
Furthermore, this technique still requires additional testing on other rock samples.

In this work, we propose a novel method to effectively represent sub-resolution solid and pore fractions using the cumulative beta distribution function, thereby improving the accuracy of stiffness computations in unresolved digital rock models.
This method requires only the measured total porosities and the digital rock images as input and is successfully validated across various rock samples, including Bentheimer sandstone, Indiana limestone, and Middle Eastern carbonate. 
By providing reliable predictions of sub-resolution fractions, this methodology opens up new possibilities for investigating the influence of the partial-volume effect. 
It enables more accurate physical property simulations in large field-of-view models reconstructed from unresolved rock images. 
This advancement is promising for enhancing the accuracy and applicability of rock elastic properties simulations when dealing with coarse and unresolved rock images.

\section{Methodology}
\subsection{Digital rock samples}
\begin{figure}
\centering
\includegraphics[width=\textwidth]{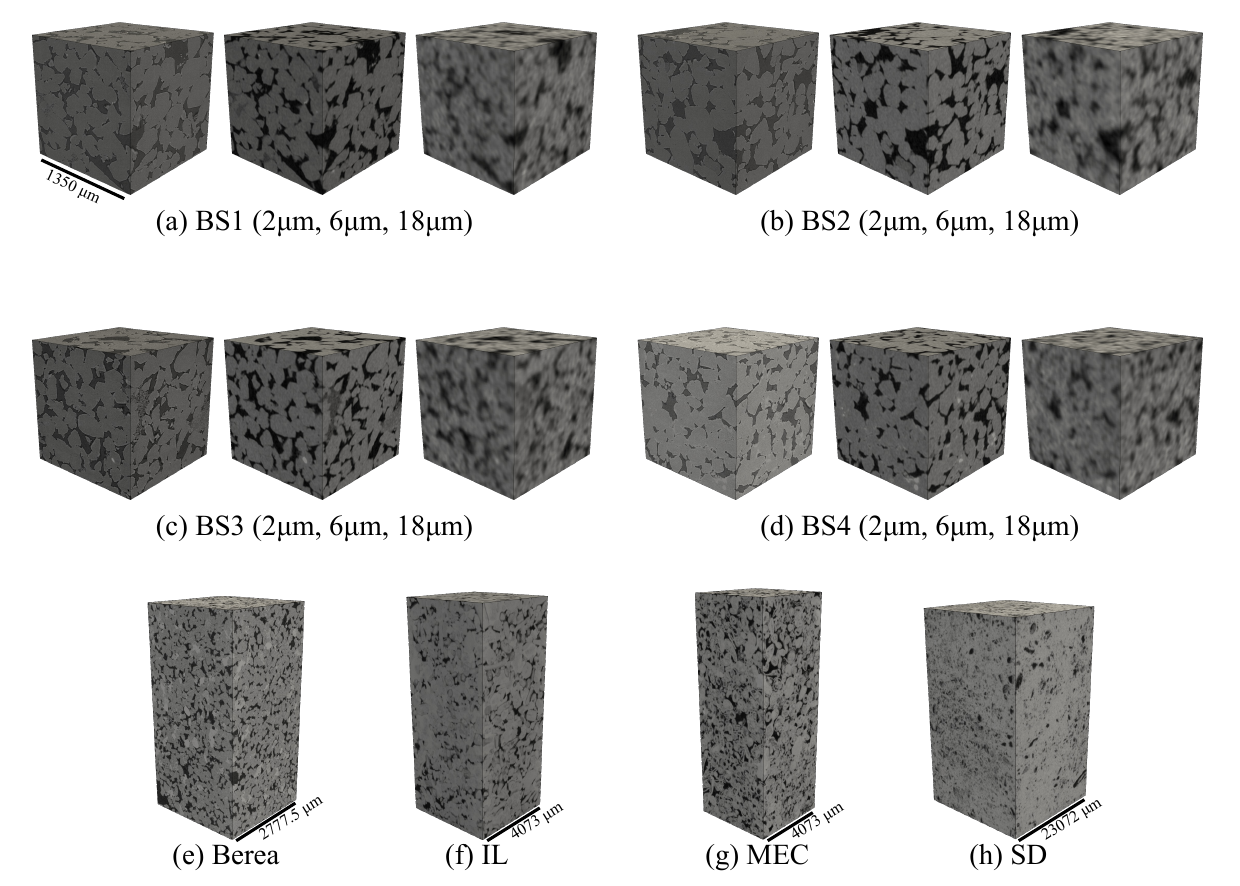}
\caption{Digital rock models used in this work. 
(a)-(d) Bentheimer Sandstone models (BS1-BS4) with resolutions of 2, 6, and 18$\mu m$;  
(e) Berea sandstone model (Berea);
(f) Indiana Limestone (IL);
(g) Middle Eastern Carbonate (MEC); 
(h) Silurian Dolomite (SD).}
\label{model figure}
\end{figure}

\begin{table}
 \caption{Digital rock models used in this work and their properties}
 \centering
 \label{Digital models}
 \begin{tabular}{p{2.5cm} cccc p{2.5cm}}
 \hline
   Rock type  &  Model & Resolution [$\mu m$] & Model size [pixels] & Porosity [\%] & Reference\\
 \hline
 \multirow{4}{*}{\makecell[l]{Bentheimer \\Sandstone}} & BS1  & \multirow{4}{*}{2, 6, 18}  & \multirow{4}{*}{$675^3$, $225^3$, $75^3$} &  22.30 & \multirow{4}{2cm}{\citet{Jackson2021}} \\
     & BS2  & & & 23.55 \\
    & BS3  &  &  & 23.32 \\
    & BS4  & &  &  19.28\\ \\
    
\makecell[l]{Berea \\Sandstone} & Berea & 4.62916 & $600\times 600\times 1200$ & 18.6 & \citet{Herring2019}\\ \\

\makecell[l]{Indiana \\Limestone} & IL & \makecell[c]{2.68,\\ 10.72} & \makecell[c]{$1000^3$, \\$380\times 380\times 888$} & \makecell[c]{9.53 (macro) \\16.38 (total)} & \multirow{4}{2cm}{\citet{Alqahtani2021}}\\ \\

\makecell[l]{Middle Eastern \\Carbonate} & MEC & \makecell[c]{2.68,\\ 10.72} & \makecell[c]{$1000^3$,\\ $380\times 380\times 1025$} & \makecell[c]{18.92 (macro) \\29.79 (total)}\\  \\

\makecell[l]{Silurian \\Dolomite} & SD & 28.84 & $800\times 800\times 1400$ & 11.8 (total) & \citet{Ferreira2020}\\
 
 \hline
 \end{tabular}
 \end{table}

In this study, we first use paired images across various rock types, including four Bentheimer sandstone  models (BS1-BS4) at resolutions of 2, 6, and $18\mu m$ \citep{Jackson2021}, Indiana limestone (IL) at resolutions of 2.68 and 10.72$\mu m$, and Middle Eastern carbonate (MEC) at resolutions of 2.68 and 10.72$\mu m$ \citep{Alqahtani2021}.
Figure \ref{model figure} illustrates 3D views of these digital rock models, and Table \ref{Digital models} provides properties of these models.
All these rock samples are primarily monomineralic. 
BS comprises approximately 95\% quartz with minor amounts of feldspar and clay minerals. 
IL is composed of 98.8\% calcite, along with minor quartz and clay minerals, featuring numerous intra-granular micro pores within the ooliths and solid grains.
MEC contains over 99\% of calcite with a significant amount of microporous ooliths, skeletal, and non-skeletal microporous grains.
For the purposes of this study, the minor occurrence of other potential minerals within these rocks is not considered.

We take models with the highest resolution as the ground truth for BS, IL, and MEC samples.
The 2$\mu m$ BS models are segmented using an automatic Otsu's segmentation method.
The generated porosity values (BS1: 22.30\%, BS2: 23.55\%, BS3: 23.32\%, and BS4: 19.28\%) are then used as reference porosities for the 6 and 18$\mu m$ paired models.
The 2.68$\mu m$ IL and MEC models are segmented using the watershed method, and the segmented models can be retrieved from \citet{Alqahtani2021}.
The high-resolution IL and MEC models have dimensions of $1000\times1000\times1000$.
The porosities of the high-resolution sub-volumes are taken as macro porosity (9.53\% for IL and 18.92\% for MEC in Table \ref{Digital models}), serving as references for seeking for macro pore fraction distribution in the 10.72$\mu m$ paired models.
For comparison, sub-volumes of dimensions $380\times380\times380$ are extracted from the 10.72$\mu m$ IL and MEC models, corresponding to the high-resolution sub-volumes of dimensions $1000\times 1000\times 1000$.
In addition, for the IL and MEC models, laboratory-measured total porosities of 16.38\% and 29.79\% are available, which will be used as references for investigating the total pore fraction distribution in the 10.72$\mu m$ unresolved rock images.

In addition to BS1-BS4, IL, and MEC models, we also explore the effective elastic properties of Berea sandstone (Berea) at a resolution of 4.62916$\mu m$ \citep{Herring2018} and Silurian dolomite (SD) at a resolution of 28.82$\mu m$ \citep{Ferreira2020}.
Similarly to the other models, Berea and SD models are assumed to be monomineralic, without considering the minor presence of other components other than the dominant mineral.
However, the Berea and SD models do not have high-resolution images for comparison and will only be used to validate our method in terms of the elastic properties calculation.

\subsection{Pore fraction profile from paired rock images}
\begin{figure}
\centering
\includegraphics[width=\textwidth]{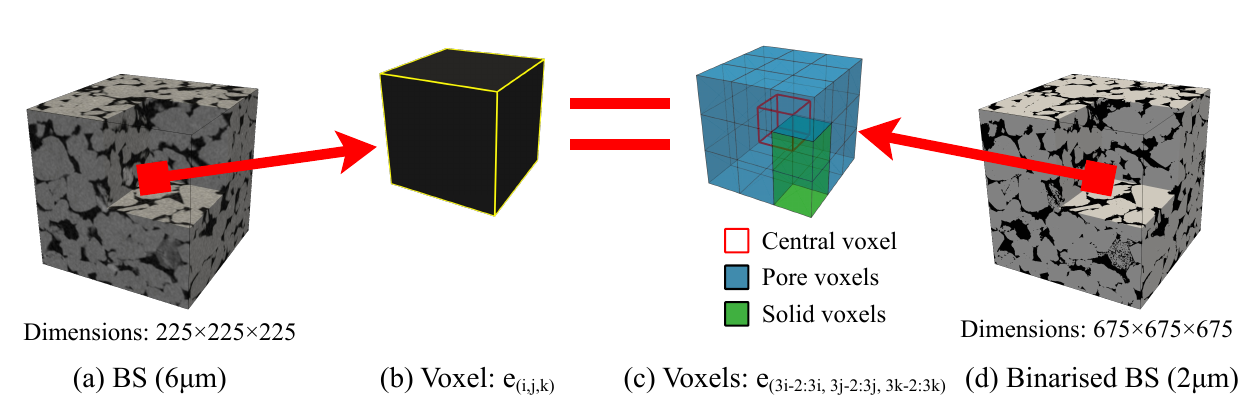}
\caption{Schematic view of the correlation of paired rock models (using BS1 as an example). 
(a) depicts the original 6$\mu m$ BS1 model; 
(b) shows a single voxel, $e_{(i,j,k)}$, extracted from (a); 
(c) is the cube containing $3\times 3\times 3$ voxels ($e_{(3i-2:3i, 3j-2:3j, 3k-2:3k)}$) extracted from the corresponding segmented 2$\mu m$ BS1 model (d). 
The voxel in (b) represents an average of the cube in (c), with the porosity of the cube in (c) considered as the actual porosity of the voxel in (b).}
\label{CT_poro}
\end{figure}

\begin{figure}
\centering
\includegraphics[width=0.6\textwidth]{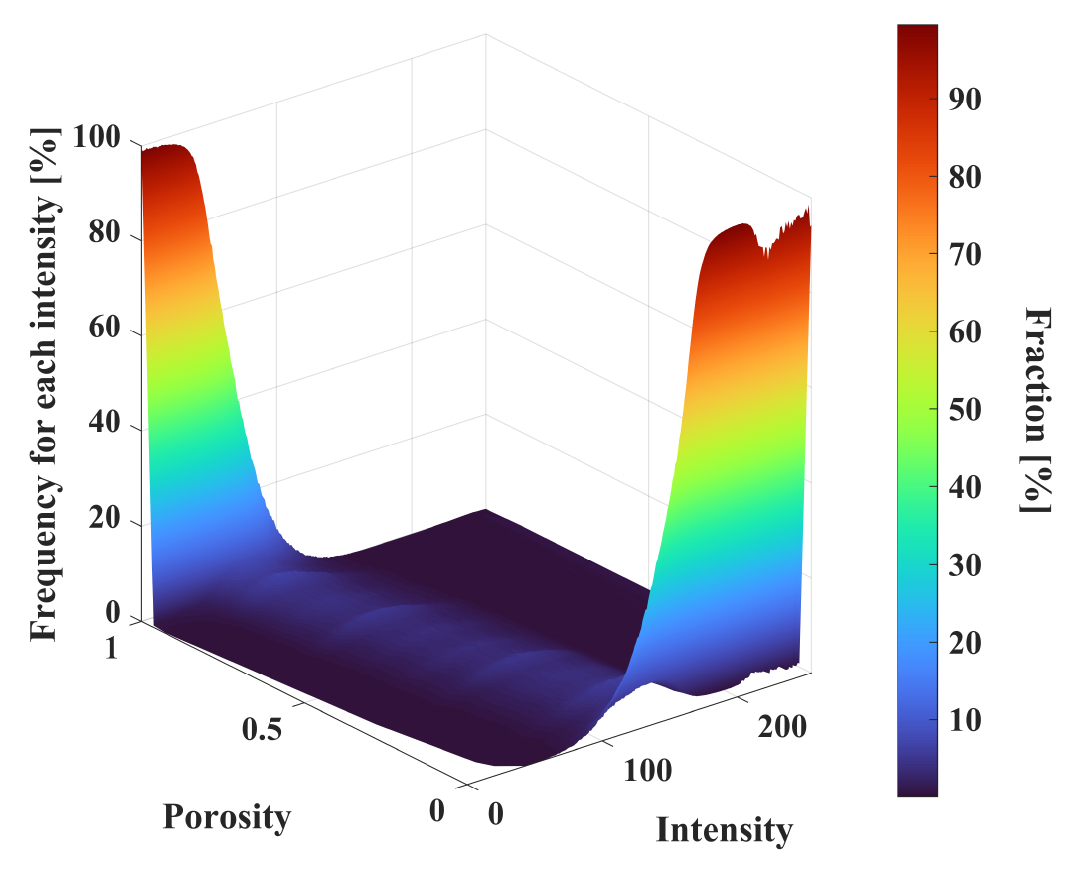}
\caption{The pore fraction distribution of voxels at each intensity in the 6$\mu m$ BS1 model, obtained by correlating the 6$\mu m$ BS1 model with its paired 2$\mu m$ model.}
\label{Porosity dist}
\end{figure}

We first determine the pore fraction distribution within unresolved BS, IL, and MEC models by comparing them with their paired high-resolution models.
In the case of the 6$\mu m$ BS model, as depicted in Figure \ref{CT_poro}, each voxel $e_{(i,j,k)}$ in this model it corresponds to an average of a cube that contains $3\times 3\times 3$ voxels ($e_{(3i-2:3i, 3j-2:3j, 3k-2:3k)}$) in its corresponding segmented 2$\mu m$ model.
The solid and pore fractions of the 6$\mu m$ voxel are therefore identical to the $3\times 3\times 3$ cube.
By counting the number of pore and solid voxels within the cube, the pore fraction of the 6$\mu m$ voxel can be easily calculated.
This process allows for the calculation of pore fraction for each voxel in the low-resolution model. 
Figure \ref{Porosity dist} illustrates the pore fractions of all voxels within the 6$\mu m$ BS1 model.
It indicates that many voxels exhibit pore fractions between 0 and 1.
Generally, voxels with higher intensity levels tend to have lower pore fractions, and vice versa. 
Interestingly, some voxels with extremely high/low intensity levels can have very high/low pore fractions, indicating the variability within voxels of the same intensity level.
Moreover, voxels containing the same pore fraction can exhibit different intensity levels.

\begin{figure}
\centering
\includegraphics[width=\textwidth]{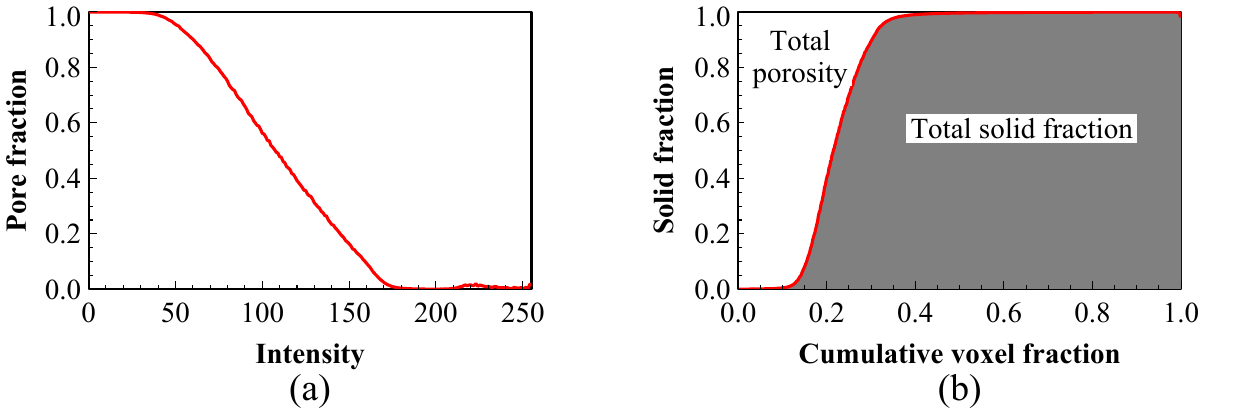}
\caption{Pore and solid fraction profiles in the 6$\mu m$ BS1 model obtained by correlating the 6$\mu m$ BS1 model with its paired 2$\mu m$ model. 
(a) shows the relation between pore fraction and intensity and (b) depicts the relationship between the solid phase fraction and the cumulative voxel frequency of the model. 
The area below the solid fraction line represents the total solid fraction of the model, while the area above it represents the total porosity.}
\label{porosity_profile}
\end{figure}

By averaging pore fraction values of voxels with the same greyscale intensity, we establish the relationship between pore fraction and greyscale intensity, as illustrated in Figure \ref{porosity_profile}(a). 
Additionally, we plot the relationship between the solid fraction ($f_s = 1 - \phi$) and the cumulative voxel frequency of the model in Figure \ref{porosity_profile}(b). 
Notably, the cumulative voxel frequency represents the total voxel volume fraction from the lowest intensity to the highest intensity, spanning from 0 to 1.
Thus, the area below the solid fraction line represents the total solid fraction, while the area above the line represents the total porosity. 
Additionally, it is evident that a significant number of voxels in the 6$\mu m$ BS1 model contain both pore and solid phases, thereby exhibiting pore/solid fractions within the range of 0-1. 
This underscores the presence of the severe partial-volume effect in unresolved rock images.


\subsection{Pore fraction profile from Beta distribution method}
Both the solid fraction and cumulative voxel frequency of the model (Figure \ref{porosity_profile}(b)) are defined in the range of [0,1]. 
The cumulative Beta distribution function can be a suitable tool for modelling the behaviour of the solid fraction profile. 
The Beta distribution function has two hyperparameters, and by adjusting these parameters, it can represent complex shapes (such as symmetric, skewed, unimodal, bimodal, etc.) within the range of 0 to 1 \citep{Johnson1995a}. 
Meanwhile, the cumulative distribution function (CDF) of the Beta distribution can represent a random monoincremental shape between 0 and 1, which is promising in capturing the relationship between the solid fraction and cumulative voxel frequency of the model, as shown in Figure \ref{porosity_profile}(b). 
This section introduces the use of the cumulative Beta distribution function to represent the solid fraction profile.

The probability density function (PDF) of a Beta distribution, $f(x; \alpha, \beta)$, is represented as follows:

\begin{equation}
	f(x\vert \alpha, \beta) = \frac{1}{B(\alpha,\beta)}x^{\alpha-1}(1-x)^{\beta-1}, \quad x = [0,1]
\end{equation}
where $\alpha$ and $\beta$ are two positive parameters that control the shape of the Beta distribution. 
$B(\alpha,\beta)$ is the Beta function, and it ensures that the total probability equals 1. 
%
%

The cumulative distribution function (CDF), denoted as $F(x\vert \alpha, \beta)$, of the Beta distribution at a given value of $x$ is expressed as:

\begin{equation}
	F(x \vert \alpha, \beta) = \frac{1}{B(\alpha,\beta)} \int_0^x t^{\alpha-1} (1-t)^{\beta-1} dt 
	\label{cdf_x}
\end{equation}

In this study, Equation \ref{cdf_x} corresponds to the solid fraction profile displayed in Figure \ref{porosity_profile}(b).
Therefore, $F(x \vert \alpha, \beta)$ is the solid fraction at any cumulative voxel frequency, $x$.
Moreover, the cumulative sum of Equation \ref{cdf_x} over the entire interval [0,1], denoted as $CF$, yields the total solid fraction within the model.
It is expressed as:

\begin{equation}
	CF = \int_0^1 F(x \vert \alpha, \beta) dx = \int_0^1 \left[\frac{1}{B(\alpha,\beta)} \int_0^x t^{\alpha-1} (1-t)^{\beta-1} dt \right]dx = f_s = 1-\phi
	\label{total cdf}
\end{equation}
where $f_s$ and $\phi$ are the total solid fraction and the total porosity of the model, respectively.

Equation \ref{total cdf} conserves the total solid fraction and porosity within the model, which can be used to determine the two parameters $\alpha$ and $\beta$ in Equation \ref{cdf_x}.
To determine the two factors, we first select two reference points $P_1$ and $P_2$ from the cumulative voxel frequency of the rock model.
$P_1$ is chosen within the pure pore phase, and $P_2$ is within the pure solid phase.
Here, we employ a Gaussian distribution process to assist in the selection of these two points.

\begin{figure}
\centering
\includegraphics[width=\textwidth]{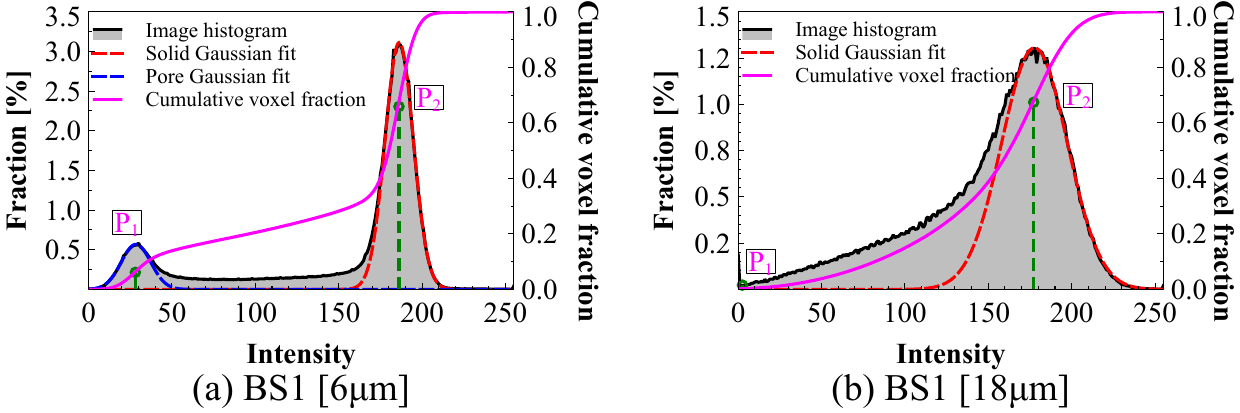}
\caption{The selection of two reference points $P_1$ and $P_2$ from the pore and solid phase Gaussian distributions and the cumulative voxel frequency. 
(a) A bimodal image intensity histogram; $P_1$ and $P_2$ are the cumulative voxel fraction at the mean intensities of the pore and solid Gaussian distributions,
(b) a unimodal image intensity histogram; $P_1$ is set as the cumulative voxel fraction at the minimum intensity in the model, while $P_2$ is the cumulative voxel fraction at the mean intensity of the solid Gaussian distribution.}
\label{Gaussian}
\end{figure}

Naturally, pure phases in digital rock images tend to exhibit a normal distribution \citep{Goldfarb2022, Ikeda2020}.
\citet{Goldfarb2022} reported that selecting the mean intensity as the pure phase intensity and treating voxels with intensities below this threshold as partial-volume voxels leads to accurate predictions of the effective elastic properties of the digital rock models.
In this work, once the parameters for the normal distributions are determined, we can plot the cumulative voxel frequency along the image intensity histogram and set the cumulative voxel frequencies at the mean intensities of the pore and solid Gaussian distributions to be $P_1$ and $P_2$, as demonstrated in Figure \ref{Gaussian}(a).
For models with low image resolutions, only the Gaussian distribution for the solid phase can be detected.
In such cases, we set $P_2$ to the cumulative voxel frequency at $\mu_s$, while $P_1$ is set as the cumulative voxel fraction at the minimum intensity in the model, as displayed in Figure \ref{Gaussian}(b).

After selecting $P_1$ and $P_2$, we substitute them into Equation \ref{cdf_x}, resulting in:
\begin{eqnarray}
	F(x = P_1 \vert \alpha, \beta) = \frac{1}{B(\alpha,\beta)} \int_0^{P_1} t^{\alpha-1} (1-t)^{\beta-1} dt < f_{s, P_1} = \frac{1}{N(P_1)} \nonumber \\ 
	F(x = P_2 \vert \alpha, \beta) = \frac{1}{B(\alpha,\beta)} \int_0^{P_2} t^{\alpha-1} (1-t)^{\beta-1} dt > f_{s,P_2} = 1-\frac{1}{N(P_2)}
	\label{constraints}
\end{eqnarray}
where $f_{s,P_1}$ and $f_{s,P_2}$ are two solid fraction criteria that constrain the values of $F(x = P_1 \vert \alpha, \beta)$ and $F(x = P_2 \vert \alpha, \beta)$; 
$N(P_1)$ and $N(P_2)$ denote the total number of voxels with intensities corresponding to the cumulative voxel fractions $P_1$ and $P_2$, respectively.

Equation \ref{constraints} stipulates that the values of $\alpha$ and $\beta$ must ensure that the solid fractions at $P_1$ and $P_2$ satisfy the specified criteria.
$F(x = P_1 \vert \alpha, \beta)<f_{s,P_1}$ imposes the constraint that there should be fewer than one solid voxel within the voxels at $P_1$, while $F(x = P_2 \vert \alpha, \beta)>f_{s,P_2}$ indicates that there should be no more than one pore voxel within voxels at $P_2$.
If both criteria are satisfied, there will be no solid voxels below $P_1$ and no pore voxel above $P_2$ in the model.
In addition, the values of $\alpha$ and $\beta$ must also satisfy the total solid fraction equation in Equation \ref{total cdf}.

We develop an algorithm to solve Equations \ref{total cdf} and \ref{constraints}.
In this algorithm, both $\alpha$ and $\beta$ iteratively vary within given wide ranges.
Among all these $\alpha$-$\beta$ pairs, we select the ones that converge the total solid fraction and porosity in Equation \ref{total cdf}.
From the selected pairs, we check if they meet the criteria defined in Equation \ref{constraints}.
If both constraints are satisfied, then the $\alpha$ and $\beta$ values are the final results.
However, in most cases, the two constraints can not be simultaneously met.
In such scenarios, one constraint is easily met while the other one is hardly achieved.
The iteration process then continues to increase the values of both $\alpha$ and $\beta$ to fulfil the unsatisfied condition, potentially resulting in excessively large $\alpha$ and $\beta$ values.
This could lead to a wide region above $P_1$ becoming pure pore phase or a wide region below $P_2$ becoming pure solid phase.
However, the objective is to utilise the $P_1$ and $P_2$ as separations of pure phases, thereby retaining the partial-volume phase between $P_1$ and $P_2$.
To address this issue, we introduce one breakpoint into the algorithm to halt the iteration process when the following condition is met:

\begin{equation}
	 \left\vert \frac{f_{s,P_2}^T-f_{s,P_2}}{f_{s,P_1}^T-f_{s,P_1}} \right\vert =
	\begin{cases}
		\geq \frac{N(P_1)}{N(P_2)} ,\quad f_{s,P_2}^T \geq f_{s,P_2} \ \& \ f_{s,P_1}^T > f_{s,P_1}\\
		\\
		\leq \frac{N(P_1)}{N(P_2)} ,\quad f_{s,P_1}^T \leq f_{s,P_1} \ \& \ f_{s,P_2}^T < f_{s,P_2} \\
	\end{cases}
\end{equation}
where $f_{s,P_1}^T$ and $f_{s,P_2}^T$ are the solid fractions at $P_1$ and $P_2$ generated by the current $\alpha$-$\beta$ pair during the iteration process.
This condition ensures that the iteration will not infinitely increase $\alpha$ and $\beta$ and produce unrealistic outcomes.

\subsection{Phase separation and elastic properties allocation}
Determining the solid fraction profile within the rock model enables us to understand the pore fraction in each voxel based on its intensity level. 
Consequently, we can segment the rock model into various phases, with each intensity representing one phase. 
We then assign appropriate bulk and shear moduli to each phase and calculate the effective elastic properties. 
However, this approach would yield too many phases that should be input into the elasticity solver, especially when dealing with 16-bit models. 
For simplicity, we develop a method to divide the rock model into a manageable number of phases while still representing the effective elastic properties.

\begin{figure}
\centering
\includegraphics[width=\textwidth]{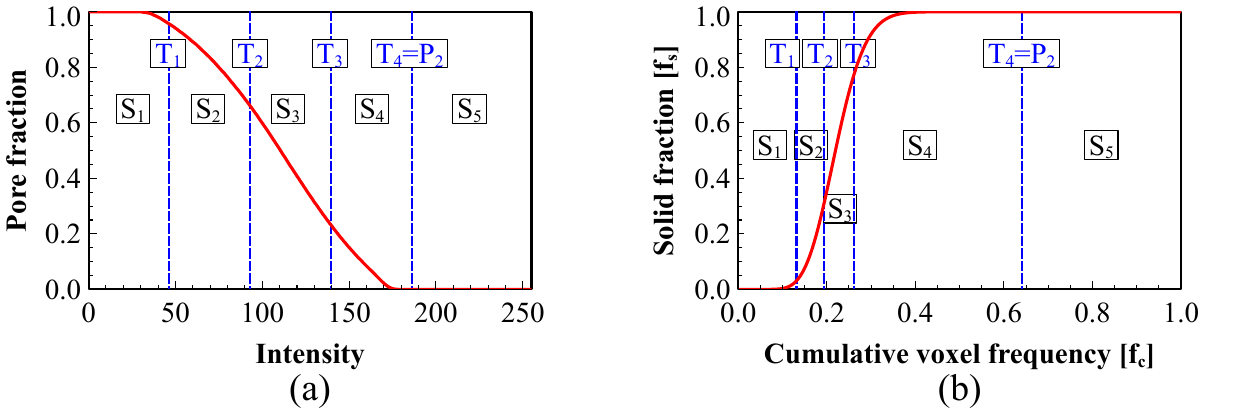}
\caption{Dividing the 6$\mu m$ BS1 model into 5 phases ($S_1$-$S_5$) by setting 4 thresholds: $T_1$-$T_4$. 
The upper limit of the threshold, $T_4$, is set to $P_2$, which is the mean intensity value of the Gaussian distribution for the solid phase.
(a) shows that $S_1$-$S_4$ cover the same intensity range.
(b) illustrates the different cumulative volume fractions covered in each phase.}
\label{separation}
\end{figure}

In this method, we evenly partition voxels below $P_2$ into $N$ phases abased on their intensity levels, treating voxels above $P_2$ as pure solid voxels.
For instance, in Figure \ref{separation}(a), we divide the entire 6$\mu m$ BS1 model into 5 phases using four thresholds ($T_1$-$T_4$). 
Four of these phases ($S_1$-$S_4$), which fall below the upper limit $T_4$ or $P_2$, are allocated the same intensity interval, while the fifth phase $S_5$ represents the pure sand phase.
Subsequently, for each sub-phase, we calculate its volume fraction and convert the intensity thresholds to cumulative volume fraction thresholds, as shown in Figure \ref{separation}(b).

We then determine the cumulative volume fraction within each sub-phase, $CF_i$, below $P_2$ using the equation:

\begin{equation}
	CF_i = \int_{T_{i-1}}^{T_{i}} F_i(x=[T_{i-1},T_{i}] \ \vert \ \alpha, \beta) dx = \int_{T_{i-1}}^{T_{i}} \left[\frac{1}{B(\alpha,\beta)} \int_0^x t^{\alpha-1} (1-t)^{\beta-1} dt \right]dx
	\label{CF_i}
\end{equation}
where $i$ ranges from 1 to $N$, and we set the threshold $T_0$ to 0 in this equation.
The pore fraction of each sub-phase, $\phi_i$, can then be computed as:

\begin{equation}
	\phi_i =1- \frac{CF_i}{T_i-T_{i-1}}
	\label{porosity of subphase}
\end{equation}

Once the pore fraction and volume fraction of each sub-phase are determined, an effective medium theory is adopted to compute their bulk and shear moduli. 
We utilise the average of the modified Hashin-Shtrikman upper and lower bounds to calculate the bulk and shear moduli of the partial-volume phases. 
The modified Hashin-Shtrikman bounds are expressed as follows:
\begin{eqnarray}
	&K_i^{\pm}&= K_1 + \frac{\phi_{i}/\phi_c}{(K_2-K_1)^{-1}+\frac{1-\phi_{i}/\phi_c}{K_1+4/3G_1}} \nonumber \\
	&G_i^{\pm}&= G_1 + \frac{\phi_{i}/\phi_c}{(G_2-G_1)^{-1}+\frac{2(1-\phi_{i}/\phi_c)(K_1+2G_1)}{5G_s(K_1+4/3G_1)}}
	\label{MHS}
\end{eqnarray}
where the superscript $+/-$ indicates the upper and lower bounds; 
$\phi_c$ is the critical porosity, which is 36\%. 
In this equation, the upper and lower bounds are calculated by swapping the materials referred to as 1 and 2. 
The upper bound is obtained when the stiffest phase is termed as material 1, whereas the lower bound is derived when the softest phase is labeled as material 1.

Equation \ref{MHS} defines the upper and lower elasticity bounds of the mixture of two materials. 
Typically, the effective elastic properties of rocks fall within these two extreme bounds. 
In sandstone, the macro pores are the predominant pore space.
Thus, the partial-volume voxels are more likely located at the boundaries of macro pores, forming contact phases between pure solid and pore phases. 
Grains are in contact with the air phase through numerous asperities in these contact phases, resulting in an intermediate overall stiffness between the stiffest and softest mixtures \citep{Goldfarb2022}. 
Therefore, the pore-solid spatial distribution in the partial-volume phase is relatively regular. 
Consequently, the effective elastic properties of the partial-volume phase are systematically lower than the upper bound. 
The bulk and shear modulus of the partial-volume phases in sandstone models are computed by averaging the upper and lower bounds:
\begin{eqnarray}
	&K_i = \frac{1}{2}(K_i^{+}+K_i^{-}) \nonumber \\
	&G_i = \frac{1}{2}(G_i^{+}+G_i^{-})
\end{eqnarray}

While carbonate rock models contain numerous micropores. 
Unlike macro pores, these micropores have sub-resolution sizes. 
Therefore, partial-volume voxels may contain complete micropores. 
In this context, the soft pore phase is surrounded by the shell of the stiff solid phase, which is consistent with the assumption of the Hashin-Shtrikman upper bound. 
Therefore, we compute the bulk and shear modulus of the partial-volume phase in carbonate models using the Hashin-Shtrikman upper bound.

\subsection{Effective elastic properties prediction}
The static Finite Element Method (FEM), as proposed by \citet{Garboczi1995}, is employed in this study to calculate effective elastic properties. 
This numerical technique is well-suited for structured grid systems and can be directly applied to models generated from digital rock images by treating each voxel as a finite element.
Consequently, it is essential to determine the bulk and shear modulus for each voxel in the FEM algorithm. 
By imposing appropriate boundary conditions, the FEM computes elastic displacements, stress, and strain within each finite element. 
This is achieved by solving a variational formulation of the linear elastic equation using an iterative algorithm, such as the conjugate gradient method \citep{Arns2002a}. 
In this work, a periodic boundary condition is applied to eliminate any boundary effects arising from additional stresses. 
Additionally, macro strains along six directions are individually applied to generate various stress distributions. 
The six macro strains will result in different stiffness components, such that we can compute the effective bulk and shear modulus of the entire model \citep{Madadi2009}.
Subsequently, the compressional and shear wave velocities of the model are calculated using formulas:

\begin{equation}
	V_p=\sqrt{\frac{K+\frac{4}{3}G}{\rho}} \hspace{0.2in} and \hspace{0.2in} V_s=\sqrt{\frac{G}{\rho}}
\label{Equation Vp_Vs}
\end{equation}
where $\rho$ is the density of the model, $V_p$ is the compressional wave velocity, $V_s$ denotes the shear wave velocity, and $K$ and $G$ are the effective bulk and shear modulus calculated by the FEM.

\section{Results}
\subsection{Solid fraction distribution}
\begin{figure}
\centering
\includegraphics[width=\textwidth]{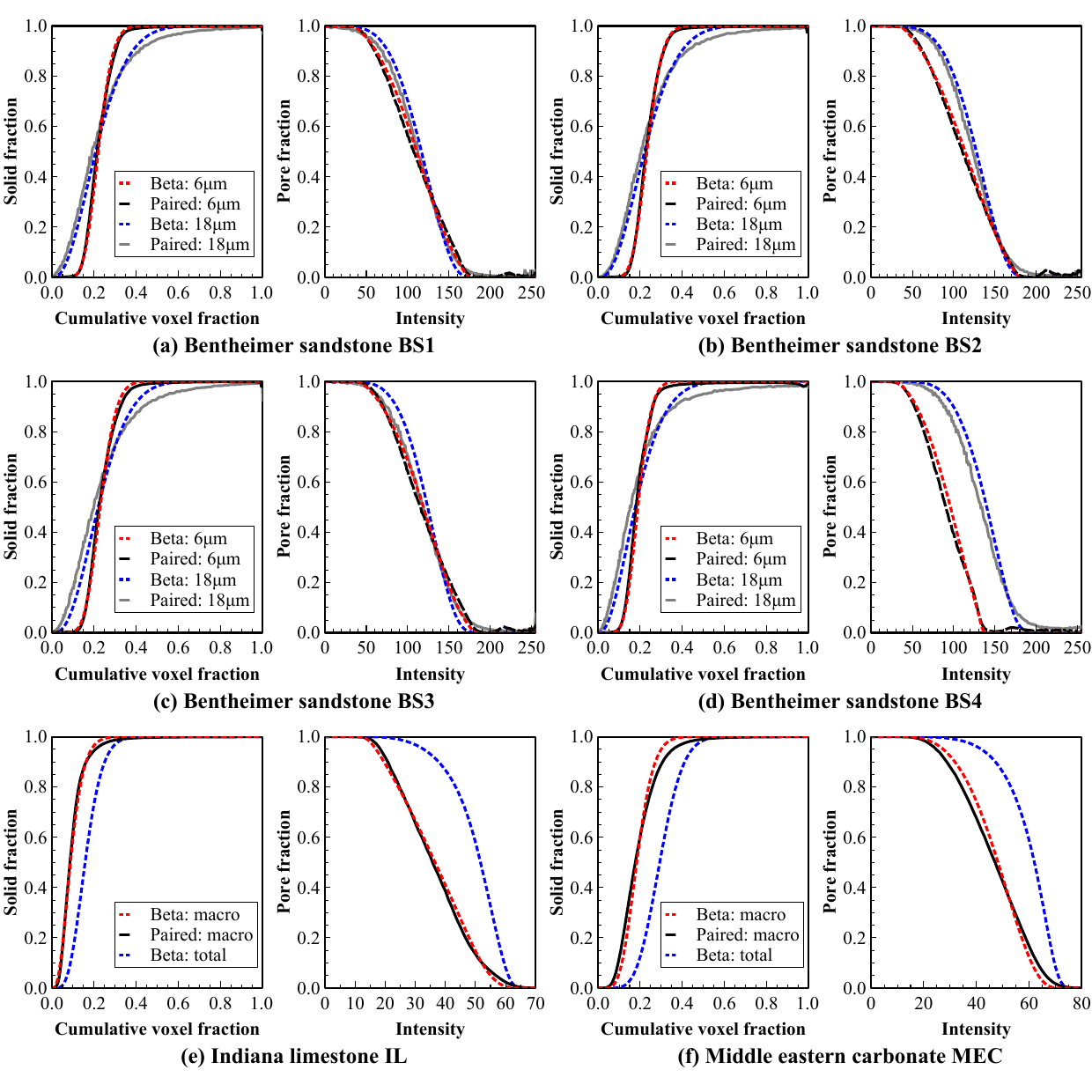}
\caption{Solid and pore fractions of different rock samples generated from the cumulative Beta distribution function ('Beta') compared to those obtained from paired images ('Paired').
Terms 'macro' and 'total' in (e) and (f) correspond to macro porosity and total porosity, respectively.}
\label{beta_match}
\end{figure}

We obtain the values of $\alpha$ and $\beta$ by executing the algorithm introduced in Section 2.3.
This allows us to determine the solid fraction profile by substituting them into Equation \ref{cdf_x}. 
To validate this method, we apply it to Bentheimer sandstone models BS1-BS4 at resolutions of 6 and 18$\mu m$, Indiana limestone model, and Middle Eastern carbonate model.
We compare the calculated solid fraction profiles with those obtained from correlating the low-resolution models to their high-resolution paired ground truth. 
Additionally, for the IL and MEC models, we use their macro porosities and total porosities in the Beta distribution method to determine the solid fraction profiles individually.

Figure \ref{beta_match} illustrates the solid and pore fractions generated by the Beta distribution method compared to those obtained from paired rock images.
The results reveal that a significant portion of voxels belong to the partial-volume phase in all tested rock models. 
In the unresolved rock models, the solid and pore fractions exhibit a broad range that varies continuously between 0 and 1.
This broad range indicates the presence of partial-volume voxels, where both pore and solid phases coexist within a single voxel due to the limited image resolution.
This emphasises the severity of the partial-volume effect present in unresolved rock images.
In addition, as the image resolution becomes coarser, the partial-volume phase region becomes broader, as evidenced by the 6 and 18$\mu m$ BS models (Figure \ref{beta_match}(a)-(d)).
The solid fraction profiles obtained from the Beta distribution method closely match those from the paired models across different rock samples.
In the 6$\mu m$ BS models, the Beta distribution method shows nearly perfect alignment with the reference.
In the 18$\mu m$ BS models, the discrepancy between the Beta distribution method and the reference becomes more pronounced.
This suggests that the Beta distribution method yields better results at higher image resolutions.
In extremely low-resolution models, the partial-volume phase widens, making it more challenging to determine the pure phase thresholds $P_1$ and $P_2$.
Arbitrary selection of these thresholds introduces additional uncertainties into the method.
Similarly, the proposed method exhibits good estimations of solid fraction distributions in the IL and MEC models.
This consistency across various rock types suggests that the cumulative Beta distribution function is capable of capturing the sub-resolution solid and pore fractions. 

\begin{figure}
\centering
\includegraphics[width=\textwidth]{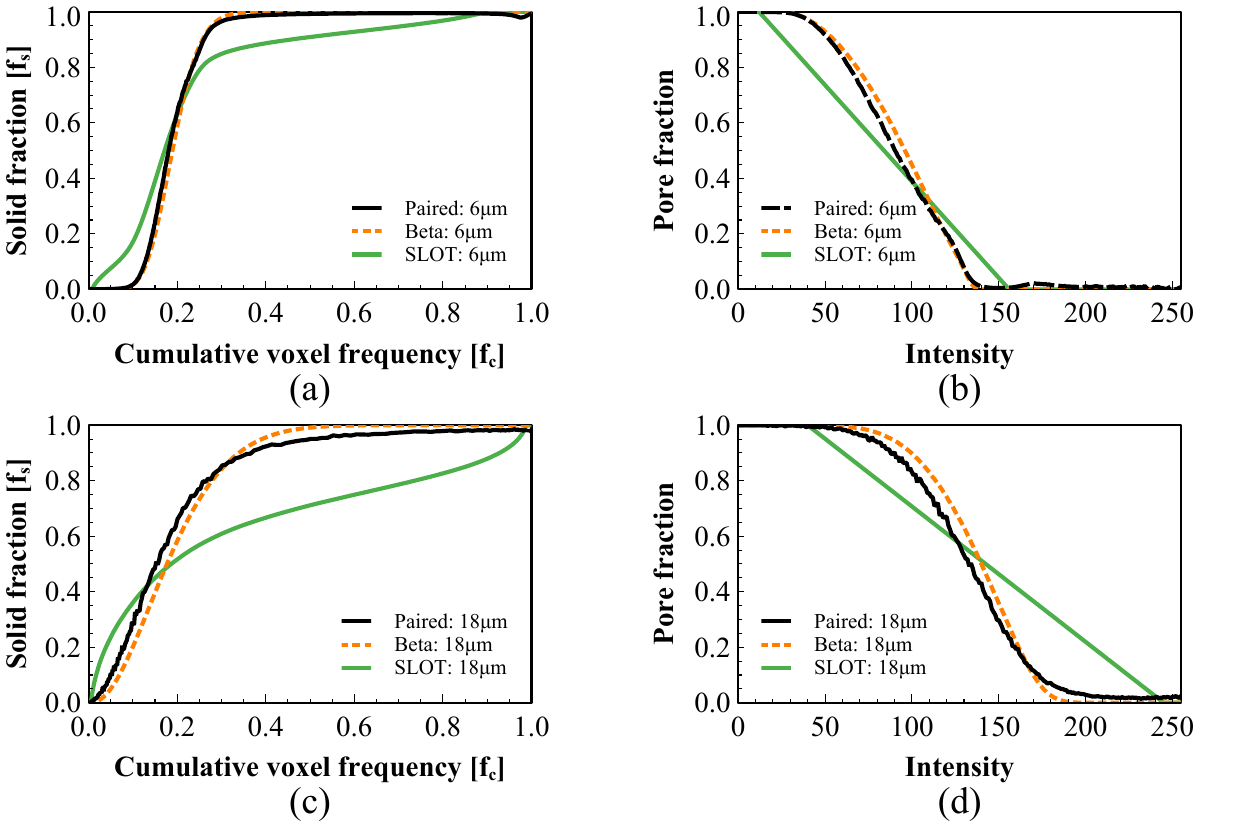}
\caption{Comparisons of solid and pore fractions of the BS4 model yielded by paired rock images (Paired), Beta distribution method (Beta), and Segmentation-less method without target (SLOT). (a) and (b) represent the 6$\mu m$ BS4 model, while (c) and (d) are the 18$\mu m$ BS4 model.}
\label{SLOT}
\end{figure}

Additionally, we implement the segmentation-less method without target (SLOT) developed by \citet{Ikeda2020} to determine solid and pore fractions in the Bentheimer sandstone models. 
This method operates through three principal steps.
First, a cubic searching window with dimensions $(2\epsilon+1)^3$ (where $\epsilon$ is a positive integer) scans the entire model to identify local maximum and minimum voxel intensities. 
Second, these extremal values are used to establish linear transformations that convert grayscale intensities to density and porosity values. 
Third, the optimal window size is determined by iteratively comparing the calculated total porosity with a reference porosity value.
The complete methodological details can be referred to \citet{Ikeda2020}.
Figure \ref{SLOT} presents a comparative analysis of solid and pore fractions obtained from three approaches: paired rock images, our Beta distribution method, and the SLOT technique. 
The results demonstrate that SLOT's linear correlation assumptions fails to represent the complex distributions of solid and pore fractions. 
This limitation becomes particularly pronounced in the $18 \mu m$ resolution model (Figures \ref{SLOT} (c) and (d)), where the discrepancies are substantially greater than those observed in the $6 \mu m$ model (Figures \ref{SLOT} (a) and (b)).

To quantitatively validate the SLOT and the Beta distribution method, we use the double-weighted mean absolute percentage error (WWMAPE), defined as:
\begin{equation}
	WWMAPE = \frac{\sum_{ct=0}^{ct_{max}} (w_{ct}|A_{ct}-A_{ct}^*|)}{\sum_{ct=0}^{ct_{max}} (w_{ct}|A_{ct}^*|)} =\frac{1}{\phi}\sum_{ct=0}^{ct_{max}} (w_{ct}|A_{ct}-A_{ct}^*|)
	\label{WWMPAE_equation}
\end{equation}
where $ct_{max}$ is the maximum greyscale intensity in the model, $w_{ct}$ is the volume fraction of voxels with intensity $ct$, $A$ and $A^*$ are the predicted and reference pore fractions.
WWMAPE incorporates data weighting, ensuring that large deviations in low-volume-fraction voxels do not significantly influence the overall error. 
Conversely, even small discrepancies in high-volume-fraction voxels can significantly impact the error. 
This weighting makes WWMAPE particularly suitable for evaluating the SLOT and Beta distribution methods.
Table \ref{Error} presents the WWMAPE values for both methods across four Bentheimer sandstone models at resolutions of 6$\mu m$ and 18$\mu m$. 
The errors increase with decreasing resolution for both methods, but the Beta distribution method consistently outperforms the SLOT. 
Specifically, the mean errors for the Beta distribution method are 3.67\% (6$\mu m$) and 13.78\% (18$\mu m$), significantly lower than those for the SLOT (24.55\% at 6$\mu m$ and 59.33\% at 18$\mu m$). 
These results demonstrate that the Beta distribution method provides more accurate pore fraction predictions by better capturing the pore fraction distribution shape compared to the SLOT's simplified linear approximation.

\begin{table}
 \caption{Double-weighted mean absolute percentage errors (WWMAPE) of SLOT and Beta distribution method across the four Bentheimer sandstone samples at resolutions of 6 and 18 $\mu m$}
 \centering
 \label{Error}
 \begin{tabular}{lccccc}
 \\
 \multicolumn{6}{l}{Resolution: $6\mu m$} \\
 \toprule
 \multirow{2}{*}{Method} & \multicolumn{5}{c}{WWMAPE}\\ \cmidrule(r){2-6}
   & BS1 [\%] & BS2 [\%] & BS3 [\%] & BS4 [\%] & Mean [\%]\\ 
   
   \midrule
 Beta & 2.98 & 2.24 & 3.71 &  5.75 & 3.67\\
 SLOT & 13.04 & 23.29 & 28.18 & 33.68 & 24.55\\
 \midrule \\

 \multicolumn{6}{l}{Resolution: $18\mu m$}\\
 \toprule
 \multirow{2}{*}{Method} & \multicolumn{5}{c}{WWMAPE}\\ \cmidrule(r){2-6}
 & BS1 [\%] & BS2 [\%] & BS3 [\%] & BS4 [\%] & Mean [\%]\\
 \midrule
 Beta & 10.68 & 10.31 & 15.42 & 18.70 & 13.78\\
 SLOT & 47.86 & 49.90 & 54.48 & 85.07 & 59.33\\ 
 \bottomrule
 \end{tabular}
\end{table}

\subsection{Segmented rock models}
\begin{figure}
\centering
\includegraphics[width=\textwidth]{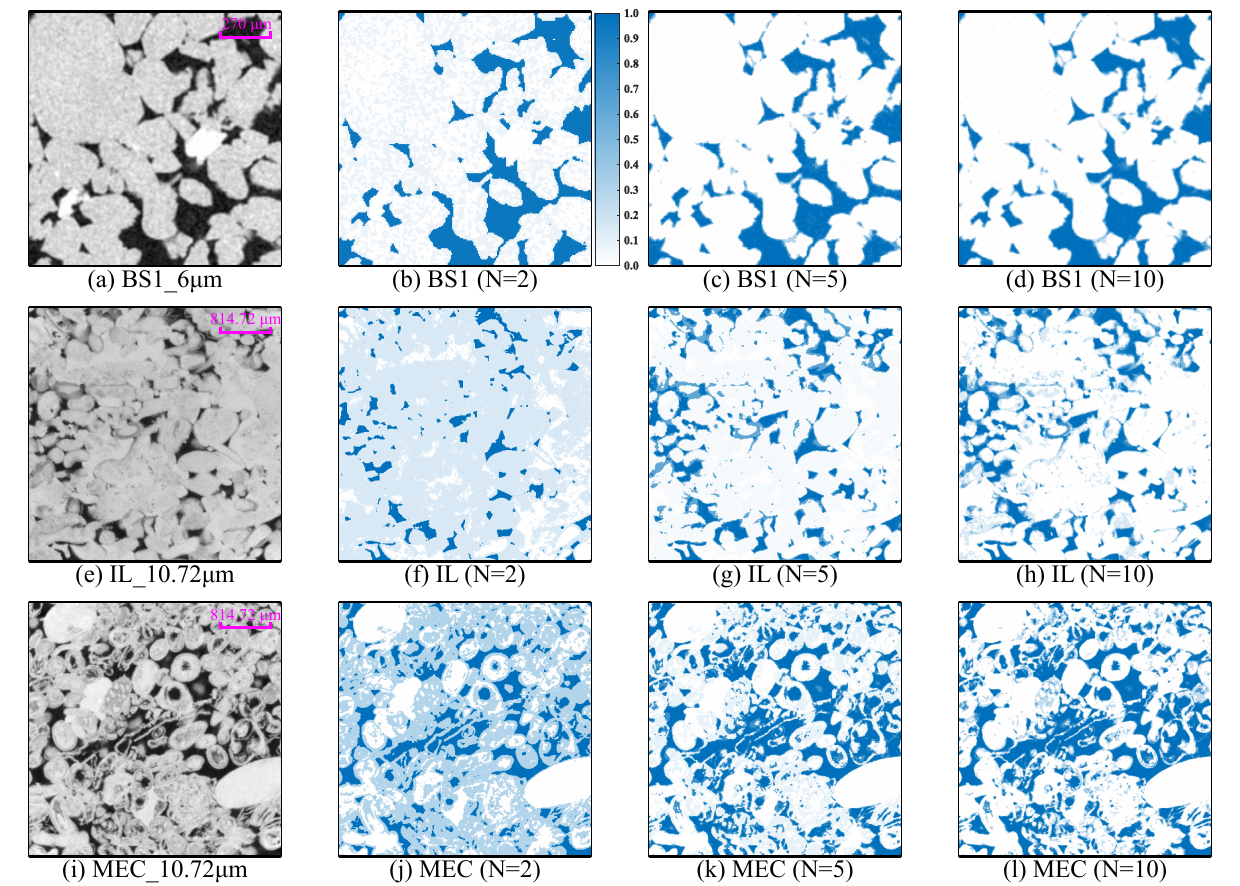}
\caption{Cross-sections depicting the pore fraction distribution within (a)-(d) BS1, (e)-(h) IL, and (i)-(l) MEC models with varying numbers of sub-phases ($N$ = 2, 5, and 10).
The left column shows the original images, while the subsequent columns (from left to right) exhibit the porosity distributions of the images with 2, 5, and 10 non-pure-solid sub-phases.}
\label{Segmentation}
\end{figure}

Figure \ref{Segmentation} illustrates the pore fraction distributions of BS1, IL, and MEC models.
Here we use the solid fraction profiles generated by the Beta distribution method (Figure \ref{beta_match}) for each rock model.
Then we vary the total number of non-pure-solid phase and calculate the pore fraction for each phase through Equations \ref{CF_i} and \ref{porosity of subphase}.
Observations indicate that increasing the number of sub-phases results in smoother transitions among phases, yielding results that closely resemble the original models.
Thus, we anticipate that as the total number of sub-phases reaches a certain value, the resultant models will become more representative of the actual models.

\begin{figure}
\centering
\includegraphics[width=\textwidth]{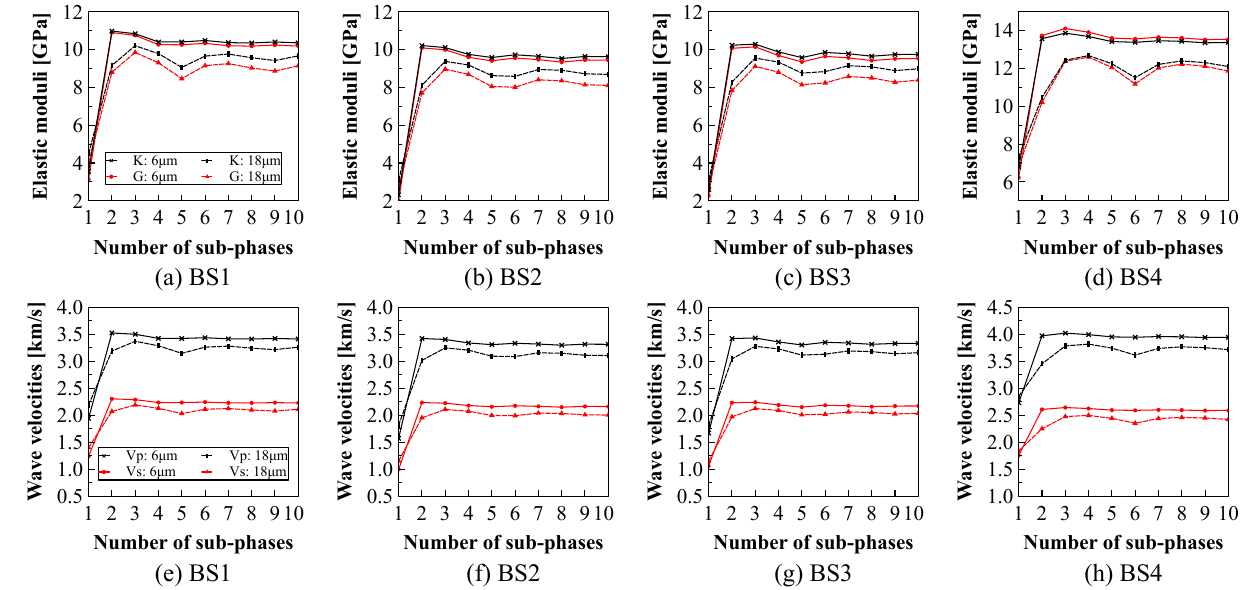}
\caption{(a)-(d) The computed bulk and shear moduli and (e)-(h) compressional and shear wave velocities of BS models (BS1-BS4) at resolutions of 6 and 18$\mu m$ with an increasing number of non-pure-solid sub-phases.}
\label{Elasticity_BS}
\end{figure}

Figure \ref{Elasticity_BS} depicts the modelled effective elastic moduli and sound wave velocities for the 6 and 18$\mu m$ BS models with varying numbers of non-pure-solid sub-phases (ranging from 1 to 10).
The properties of mineral compositions used in the simulations are provided in Table \ref{Mineral}.
It is evident that the effective elastic properties stabilise as the number of macro-mixed sub-phases increases.
The stabilised values are thus considered as the final elastic moduli and wave velocities.
Here, for the BS models, we adopt $N=10$ as the final results.

\begin{table}
 \caption{Density, bulk and shear moduli of mineral compositions used in FEM simulations.}
 \centering
 \label{Mineral}
 \begin{tabular}{lccc}
 \toprule
   Mineral &  Density [$g/cm^3$] & Bulk modulus [GPa] & Shear modulus [GPa] \\ \midrule
 Pore & 0 & 0 & 0 \\
Quartz & 2650 & 37 & 44 \\
Calcite & 2710 & 70.2 & 29 \\
Dolomite & 2870 & 76.4 & 49.7 \\ 
\bottomrule
\multicolumn{4}{l}{\textit{Note}. All properties are obtained from \citet{Mavko2009}}
 \end{tabular}
 \end{table}

\subsection{Elastic moduli and sound wave velocities}
\begin{figure}
\centering
\includegraphics[width=\textwidth]{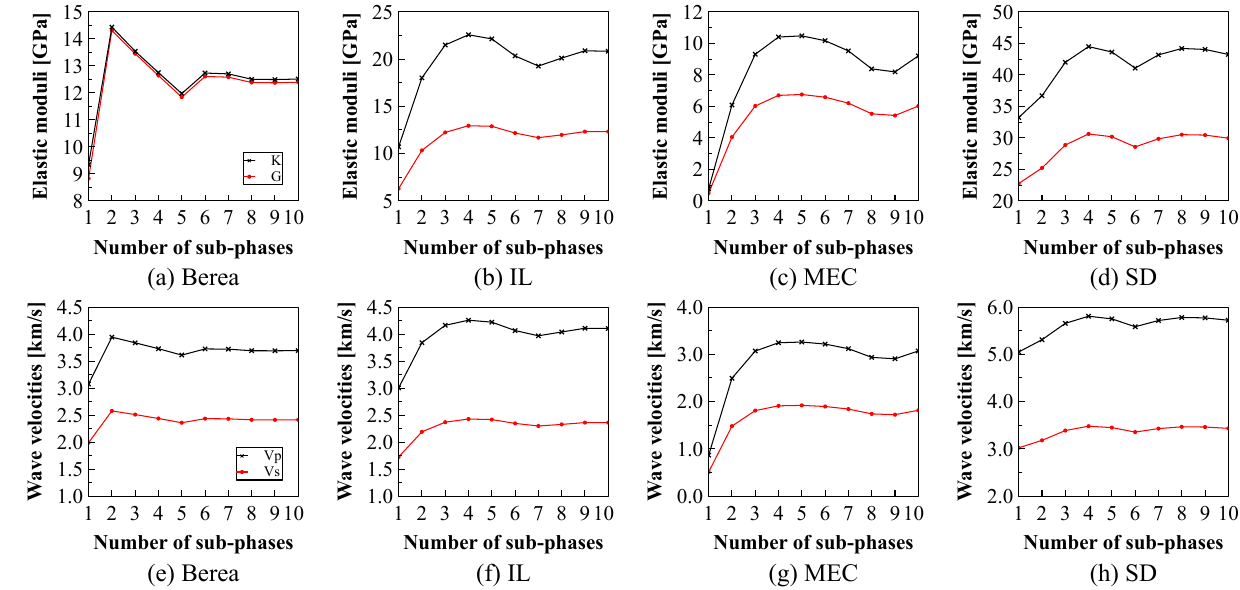}
\caption{(a)-(d) The computed bulk and shear moduli and (e)-(h) compressional and shear wave velocities of the Berea, IL, MEC, and SD sub-volumes with an increasing number of sub-phases.}
\label{Elasticity_others}
\end{figure}

For other models including Berea, IL, MEC, and SD, we first obtain their solid fraction profiles using the Beta distribution method.
We then systematically vary the number of sub-phases to ascertain their stable elastic properties.
However, due to their large Z-axis dimensions as full core models, we partition them into smaller sub-volumes before segmenting and calculating their effective elastic properties.  
Specifically, the original $600\times 600\times 1200$ Berea model is subdivided into 24 sub-models of dimensions $600 \times 600 \times 50$. 
Likewise, the IL model, with dimensions $380 \times 380 \times 888$, is divided into 24 smaller models of dimensions $380\times 380 \times 37$. 
Similarly, the MEC model, with dimensions $380\times 380 \times 1025$, is divided into 25 models of dimensions $380 \times 380 \times 41$. 
Lastly, the SD model of dimensions $800\times 800 \times 1400$, is segmented into 25 sub-volumes of dimensions $800\times 800\times 50$.
For each type of rock model, we select one sub-volume and compute its effective elastic properties while incrementally varying the number of sub-phases. 
The resulting elastic properties are depicted in Figure \ref{Elasticity_others}. 
We consider the values obtained when $N=10$ as the final elastic properties. 
Subsequently, for the remaining sub-volumes of each rock type, we directly divide the non-pure-solid phase into 10 sub-phases and calculate their effective elastic properties accordingly.

\begin{figure}
\centering
\includegraphics[width=\textwidth]{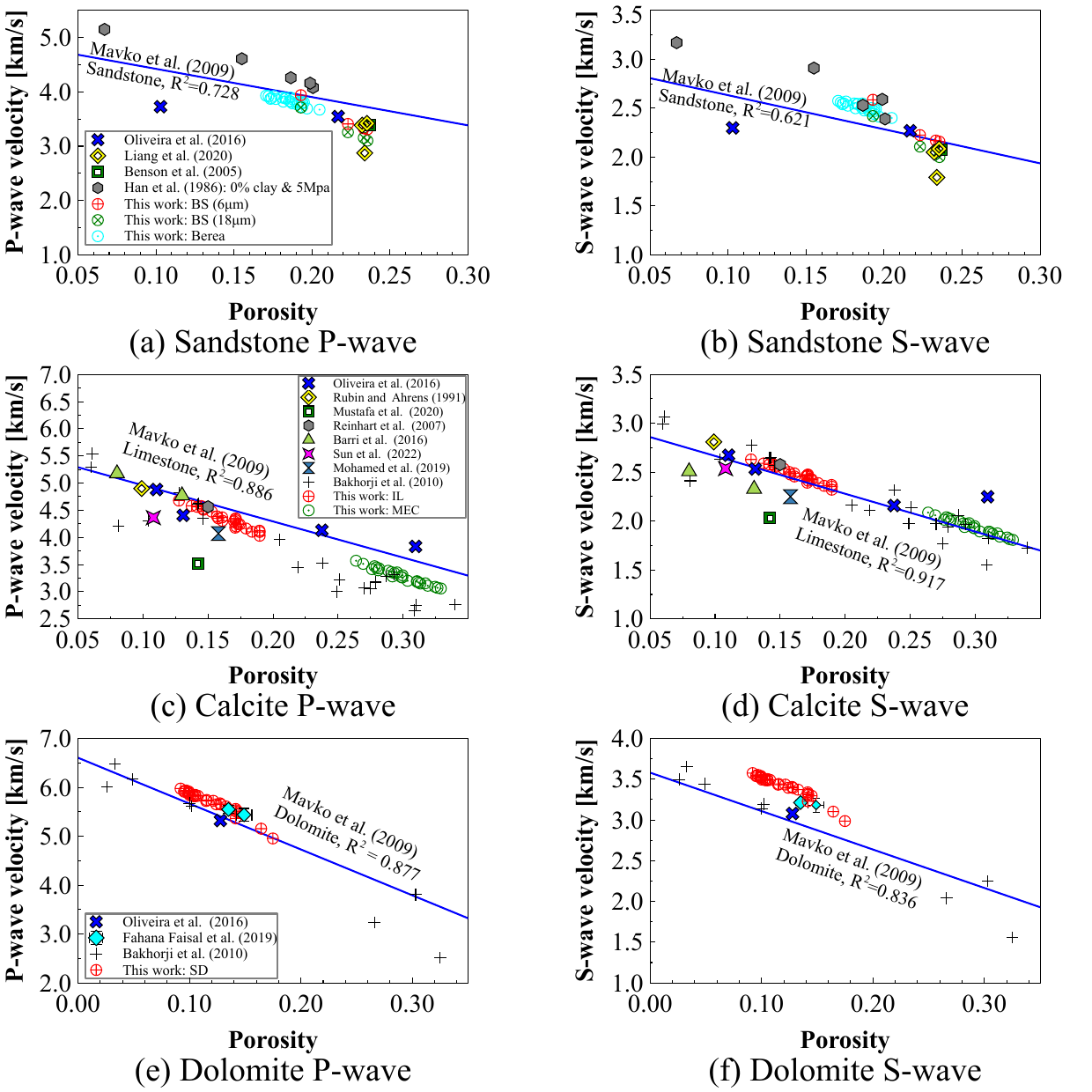}
\caption{The comparison of compressional and shear wave velocities between the modelled results and some documented experiment measurements for different rock types.
The left column shows the P-wave velocities of (a) sandstone, (c) calcite carbonate, and (e) dolomite;
While the right column displays the S-wave velocities of (b) sandstone, (d) calcite, and (f) dolomite, respectively.
The experimental results are taken from: (1) sandstone: \citet{Oliveira2016}, \citet{Liang2020}, \citet{Benson2005}, and \citet{Han1986}; (2) calcite: \citet{Oliveira2016}, \citet{Rubin1991}, \citet{Mustafa2020}, \citet{Reinhart2008}, \citet{Barri2016}, \citet{Sun2022}, \citet{Mohamed2019a}, and \citet{Bakhorji2010}; (3) dolomite: \citet{Oliveira2016}, \citet{FarhanaFaisal2019} , and \citet{Bakhorji2010}.
The trend lines in these figures represent the regressions of the wave velocities and the porosity of the corresponding rock types, taken from \citet{Mavko2009}.
}
\label{Wave comparisons}
\end{figure}

Figure \ref{Wave comparisons} illustrates the comparison between the modelled compressional and shear wave velocities and documented experimental measurements for different rock types.
It is important to note that the experimental results presented in Figure \ref{Wave comparisons} were obtained under zero or very low ($\leq$5MPa) confining pressure.
Therefore, the rocks are expected to still contain some micro-pore features that will close under elevated confining pressure.
These micro-pore features are present under low confining pressure, which can drastically affect the effective elastic properties of the sample.
The good agreement between the modelled results and experimental measurements demonstrates that our method can effectively predict the elastic properties of rocks under very low confining pressure conditions.

\section{Discussion}
We have demonstrated that the Beta distribution method is effective in revealing the pore/solid fraction distribution within rock models constructed from unresolved digital images.
Incorporating these pore/solid fraction distributions into the reconstruction of digital rock models can improve the accuracy of elastic property simulations.
Therefore, this method holds promise in areas that the quality of estimating rock elastic properties is crucial, such as reservoir characterisation, well logging interpretation, and digital rock core analysis.
Additionally, it can be particularly helpful in conducting simulations in large-scale digital rock models constructed from low-resolution images.

However, some important considerations must be taken into account when applying this method. 
Firstly, the assumption that higher intensity represents higher solid fraction and lower pore fraction may not always hold true. 
The pore fraction calculated by this method represents an average pore fraction at different intensity levels, but it does not account for variations within voxels of the same intensity. 
High-pore-fraction voxels with high intensity levels may indicate the presence of micro-cracks surrounded by high-intensity solid compositions, which can significantly affect the accuracy of the effective elastic property simulation. 

Secondly, this method is applicable primarily to monomineralic or nearly monomineralic rock samples, where no other mineral phases exist between the dominant solid phase and the pore phase. 
Additionally, measured porosity plays a crucial role in this method as it directly influences the resultant porosity profiles in rock models. 

Lastly, the effective medium theory used in this work plays a vital role in the calculation of effective elastic properties of rock models. 
We used the average of the modified Hashin-Strikman bounds for sandstone models, while the upper bound is used for carbonate rock models according to their pore-solid spatial distributions. 
However, in both macropore-dominant rocks, such as sandstones, or micropore-dominant rocks, such as carbonate rocks, the image resolution affects the pore-solid spatial distribution within their partial-volume voxels. 
Coarsening of the image resolution makes it more possible to cover complete pores within partial-volume voxels. 
Therefore, the effective elastic properties of partial-volume phases should be closer to the MHS upper bound with the decrease in image resolution. 
Thus, future work is encouraged to find more suitable theoretical models to calculate the bulk and shear modulus for partial-volume voxels.

\section{Conclusions}
In this study, we proposed a novel method to reconstruct digital rock models from unresolved rock images with improved elastic moduli and sound wave velocity properties. 
This method uses the cumulative Beta distribution function to effectively represent the sub-resolution pore fraction in unresolved rock images.
We systematically compared the pore fraction distribution within partial-volume voxels in various rock samples using the proposed method against that from paired rock images scanned at multiple resolutions. 
Additionally, we compared the compressional and shear wave velocities of rock models reconstructed from the proposed method to documented experimental results. 
The results indicate that a high volume fraction of pore space can be contained in the partial-volume phase, and the quantity of the partial-volume phase increases with the coarsening of image resolution.
The proposed method adopts the cumulative Beta distribution function, effectively revealing the relationship between image intensity and pore/solid fraction within each voxel in various rock samples and image resolutions. 
The double-weighted mean absolute percentage errors of predicted pore fractions for the proposed method are 3.67\% (6$\mu m$) and 13.78\% (18$\mu m$) across the four tested Bentheimer sandstone samples, which is significantly lower than the SLOT technique's errors of 24.55\% (6$\mu m$) and 59.33\% (18$\mu m$).
Furthermore, the modelled sound wave velocities of different rock models generated by the proposed method agree well with documented experimental results. 
This demonstrates the capability of the proposed method in unraveling the pore fraction distribution and reconstructing more accurate digital rock models from unresolved rock images. 
Thus, it shows promise in facilitating simulations within large-scale unresolved rock models.

\section*{Declaration of Competing Interest}
The authors declare that they have no known competing financial interests or personal relationships that could have appeared to influence the work reported in this paper.

\section*{Acknowledgement}
The financial support for this study was provided by PetroChina.

\section{Open Research}
The code used for modelling sub-resolution solid and pore fractions is available at \url{https://doi.org/10.5281/zenodo.13239406}.

The Bentheimer sandstone images are available at \citet{Jackson2021} \\ (\url{https://zenodo.org/record/5542624}).

The Berea sandstone images are from \citet{Herring2019}, which can be accessed from \citet{Herring2018} (\url{https://www.digitalrocksportal.org/projects/135}).

The Indiana limestone and Middle eastern carbonate images are from \citet{Alqahtani2021} (\url{https://www.digitalrocksportal.org/projects/362}).

The Silurian dolomite images are from \citet{Ferreira2020}
\\ (\url{https://www.digitalrocksportal.org/projects/252}).

%



  \bibliographystyle{elsarticle-harv} 
  \bibliography{Reference}



%
%
%

\end{document}